\documentclass[aps,prb,twocolumn,superscriptaddress]{revtex4}
\usepackage{color}
\usepackage{amsmath}
\usepackage{graphicx}
\usepackage{tabularx}
\usepackage{textcomp}
\usepackage{gensymb}
\usepackage{keyval}
\usepackage{multirow}
\usepackage[normalem]{ulem}
\newcommand{\eye}{\mathrm{i}}
\newcommand{\pdag}{{\phantom\dagger}}
\renewcommand{\vec}[1]{\boldsymbol{#1}}

\begin{document}
\preprint{0}

\title{From bad metal to Kondo insulator: temperature evolution of the optical properties of SmB$_{6}$}

\author{A. Tytarenko}
\affiliation{Van der Waals-Zeeman Institute, Institute of Physics (IoP), University of Amsterdam, Science Park 904, 1098 XH, Amsterdam, the Netherlands}
\author{K. Nakatsukasa}
\affiliation{Department of Physics and Astronomy, University of Tennessee, Knoxville, Tennessee 37996, U.S.A}
\author{Y.K. Huang}
\affiliation{Van der Waals-Zeeman Institute, Institute of Physics (IoP), University of Amsterdam, Science Park 904, 1098 XH, Amsterdam, the Netherlands}
\author{S. Johnston}
\affiliation{Department of Physics and Astronomy, University of Tennessee, Knoxville, Tennessee 37996, U.S.A}
\author{E. van Heumen}
\email{e.vanheumen@uva.nl} 
\affiliation{Van der Waals-Zeeman Institute, Institute of Physics (IoP), University of Amsterdam, Science Park 904, 1098 XH, Amsterdam, the Netherlands}

\begin{abstract}
The recent rekindling of interest in the mixed valent Kondo insulator SmB$_{6}$ as candidate for a first correlated topological insulator has resulted in a wealth of new experimental observations. In particular, angle-resolved photoemission experiments have provided completely new insights into the formation of the low temperature Kondo insulating state starting from the high temperature correlated metal. Here, we report detailed temperature and energy dependent measurements of the optical constants of SmB$_6$ in order to provide a detailed study from the point of view of a bulk sensitive spectroscopic probe. We detect a previously unobserved infrared active optical phonon mode, involving the movement of the Sm ions against the boron cages. The changes taking place in the free carrier response with temperature and their connection to changes in optical transitions between different bands are discussed. We find that the free charge density starts to decrease rapidly below approximately 200 K. Below 60 K a small amount of spectral weight begins to accumulate in low lying interband transitions, indicating the formation of the Kondo insulating state; however, the total integrated spectral weight in our experimental window ($\sim 4.35$ eV) decreases. This  indicates the involvement of a large Coulomb interaction ($>$ 5 eV) in the formation of the Kondo insulator.
\end{abstract}

\maketitle
\section{Introduction}
The enormous interest in SmB$_{6}$ as the first realization of a correlated topological insulator (or in this case more precisely a topological Kondo insulator \cite{Dzero:2010dj,Dzero:2015dza}), has resulted in a wealth of new information on its electronic properties and the formation of the Kondo insulating state. In a concerted effort to detect the predicted surface states, several angle-resolved photoemission spectroscopy (ARPES) experiments \cite{Jiang:2013hj, Zhu:2013fg, Denlinger:2013tk, Frantzeskakis:2013vy, Xu:2014im, Xu:1jv, Min:2014kp, Denlinger:2014bj}, transport experiments \cite{Wolgast:2013ih, Kim:2013gq, Zhang:2013em, Syers:2015da, Luo:2015fe, Chen:2015id, Tan:2015wf}, and scanning tunneling microscopy (STM) experiments \cite{Ruan:2014db, Rossler:2014kn} have been performed.  At the same time, fine tuning of the crystal growth conditions as well as comparisons between growth methods have been reported\cite{Hatnean:2013kta,Bao:2013iw,Phelan:2016kg}, making it worthwhile to also revisit bulk sensitive probes such as optical spectroscopy. Previous optical studies \cite{KierzekPecold:1969tu, Allen:1978ii, Gorshunov:1999hs, Sluchanko:2000hf,Nanba:2002va, Hudakova:2004bh, Travalgilini:1984vy} have provided important information on the formation of the Kondo insulating state.  

In this article we report reflectivity measurements of SmB$_6$, and present an analysis of the optical conductivity and dielectric function derived from them.  As the temperature below which surface states are expected to become detectable is below the lowest temperature achievable in our setup, we instead focus on the changes in the optical spectra related to the formation of the Kondo insulating state. The room temperature spectra are characterized by an incoherent metallic response and several interband transitions. The interband transitions can be related to LSDA+$U$ calculations \cite{Antonov:2002uw},  where Sm has been taken to have both a 2$^+$ and 3$^+$ valence, providing an indirect signature of the mixed valence state of Sm. We also observe for the first time an infrared mode at 19.4 meV that we associate with the motion of Sm ions against the B$_{6}$ cages. As temperature is decreased, the incoherent metallic response first becomes more coherent but then collapses below approximately 60 -- 70 K. The total amount of spectral weight lost in the metallic response corresponds to approximately 0.076 carriers per SmB$_{6}$ unit and is not recovered in the entire energy range of the experiment (4.35 eV). Based on a simple calculation we rule out kinematic effects associated with the hybridization gap as the cause for the lost spectral weight. Instead, we suggest that strong correlation effects, associated with the effective Coulomb interaction $U$, are responsible for the transfer of spectral weight to very high energies that has thus far not been observed in mixed valent Kondo insulators.

The paper is organized as follows: in Sec. \ref{exp_sec} we discuss the experimental details and present an overview of the main optical features that can be directly inferred from the reflectivity data. In Sec. \ref{opt_cond}, we present the optical response functions and present an identification of the main features. A detailed analysis of the transfer of spectral weight is presented in Sec. \ref{spec_transf}. 
(Our toy model calculations are presented in Appendix \ref{app_tb}.)
Finally, in Sec. \ref{summary} we provide a brief summary and concluding remarks.

\section{Methods.}\label{exp_sec}
A single crystal boule was grown by the floating-zone technique, as detailed previously in Ref. [\onlinecite{Bao:2013iw}]. To obtain a large mirror-like surface, we first oriented the as-grown boule using Laue diffraction. We then made a slit using spark erosion, cutting on the side of the crystal along the (100) oriented plane and subsequently cleaved the crystal just before inserting it in our UHV cryostat. This resulted in a large (5 mm diameter) flat surface oriented perpendicular to the [001] direction. 

The reflectivity of SmB$_6$ was measured as a function of photon energy (4 meV -- 4.35 eV) and temperature (14 -- 300 K) with 2 K steps. The measurements were performed using a Bruker vertex 80v Fourier-Transform infrared spectrometer. The crystal was mounted in a high-vacuum cryostat ($\sim 10^{-9}$ mbar) that includes \textit{in - situ} evaporators to cover the sample with a reference layer. Different materials (e.g. Au, Ag, Al) have been used as reference layers for different frequency ranges. A detailed description of the experimental procedures, which we used to calibrate the obtained reflectivity, are given in Ref. [\onlinecite{Tytarenko:2015jj}]. 

\begin{figure}
 \includegraphics[width = 8.6 cm]{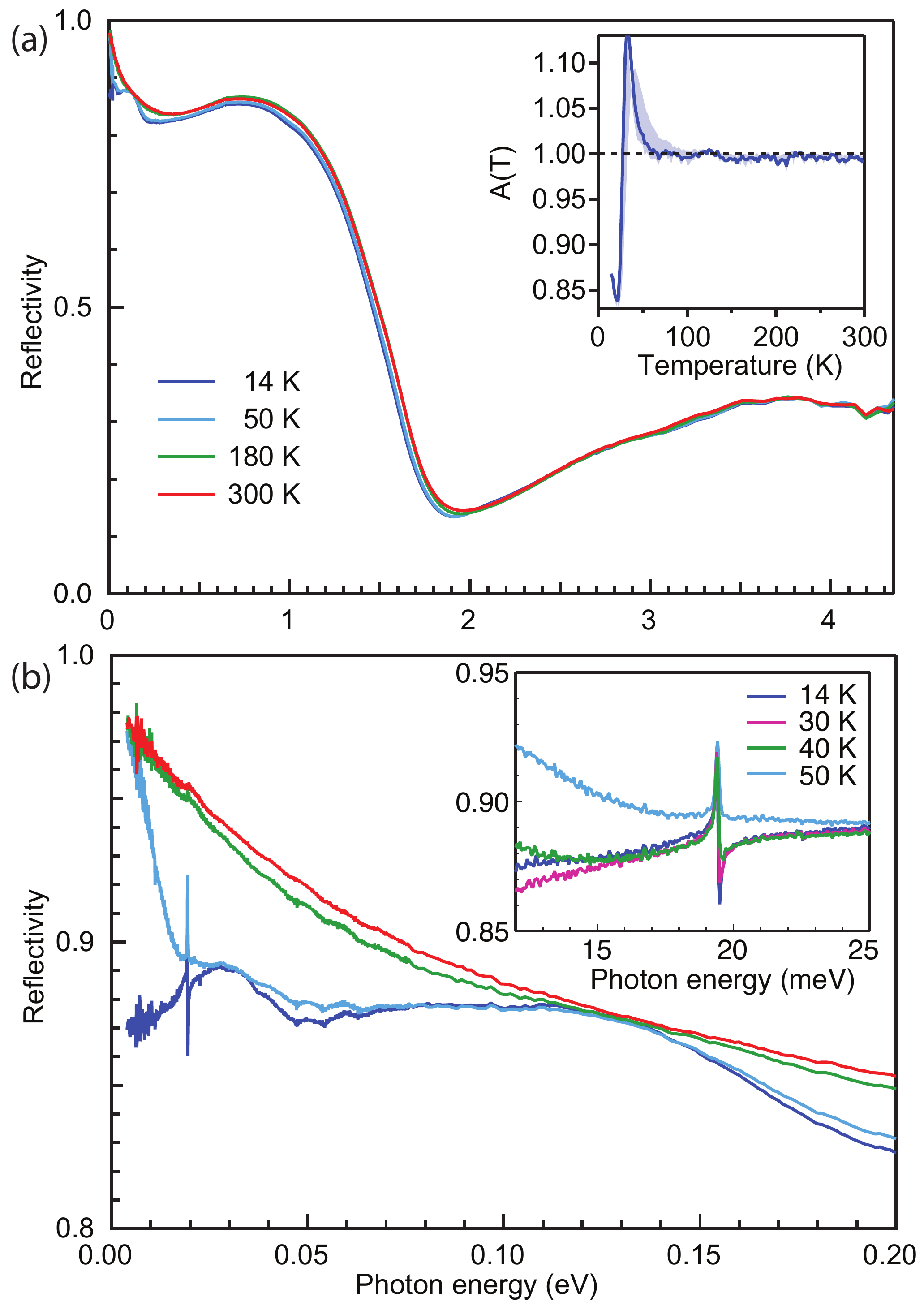}
 \caption{(a): The reflectivity of SmB$_6$ at selected temperatures over the full frequency range measured. A small shift in the plasma edge is visible, as well as temperature dependent changes in the UV range. The inset shows $A(T)$ (discussed in the text). The deviation of $A(T)$ from 1 indicates the temperature below which Hagen-Rubens (HR) behavior is lost. The light shaded region indicates the range of values obtained for $A(T)$ by changing the fitted frequency range.  (b): Low frequency reflectivity  at the same temperatures as in panel (a). With decreasing temperature a minimum develops below the isobestic point around 0.13 eV. At the lowest measured temperature (14 K) the reflectivity spectra resembles that of a semiconductor. The sharp line visible at low temperature around 20 meV is the $T_\mathrm{1u}$ vibrational mode, which is shown in more detail and at different temperatures in the inset.}
\label{Fig1}
\end{figure}

Reflectivity spectra for selected temperatures are shown in Fig. \ref{Fig1}.  Figure \ref{Fig1}a shows the reflectivity over the full spectral range measured. The most prominent changes as function of temperature take place in the low frequency part of the spectrum (displayed in Fig. \ref{Fig1}b).  However, at higher photon energies a clear shift of the plasma edge is visible between 1 and 2 eV. Above the plasma edge several structures in the reflectivity point to interband transitions. Figure \ref{Fig1}b shows the reflectivity below 0.2 eV, where the effects due to the formation of the Kondo state at low temperature are most prominent. Overall the spectral features agree with previously published data \cite{Travalgilini:1984vy, Nanba:2002va, Gorshunov:1999hs, Hudakova:2004bh}, but there are also significant differences as we will discuss further below. At room temperature the reflectivity scales according to the expectation for a metal, namely as $R(\omega) \approx1-2\sqrt{\omega / \sigma_\mathrm{DC}}$ (i.e. Hagen - Rubens behavior), with $\sigma_\mathrm{DC}$ the DC conductivity. We extracted $\sigma_\mathrm{DC}$ by fitting the Hagen-Rubens relation to the low energy $R(\omega)$. It agrees to within a factor 2-3 with the DC resistivity measured on a different crystal taken from the same boule (data reported in Ref. [\onlinecite{Frantzeskakis:2013vy}]). 

At room temperature, Hagen-Rubens (HR) behavior persists up to about 12 meV, but we expect the HR scaling to eventually break down below a certain temperature due to the opening of the hybridization gap. We estimate this temperature scale as follows: we fit the low frequency reflectivity to the relation $R(\omega,T)=A(T)-B(T)\sqrt{\omega}$ and determine $A(T)$ and $B(T)$.  The functions $A(T)= 1$ and $B(T)=\sqrt{4/\sigma_\mathrm{DC}}$ in the Hagen-Rubens limit.  From $B(T)$ we obtain $\rho_\mathrm{DC}(T)$, while a deviation of $A(T)$ from 1 is taken as an indication for the opening of the hybridization gap. We can directly compare $\rho_\mathrm{DC}(T)$ obtained from our fits to the temperature dependent resistivity reported in Ref. [\onlinecite{Frantzeskakis:2013vy}]. Apart from a (temperature independent) scaling factor $\approx$ 2.7, we obtain excellent agreement between these two measurements if we restrict the photon energy range used in the HR fit from the lower limit of the data (4 meV) up to 12 meV. The temperature dependence of $A(T)$ obtained in this case is shown in the inset of Fig. \ref{Fig1}a. The blue shaded area indicates the dependence of $A(T)$ resulting from varying the upper bound of the photon energy range (6 meV -- 30 meV) included in the fits. Based on the departure of $A(T)$ from one, we estimate that the metallic response persists down to approximately 60 - 70 K, which is in good agreement with estimates of the onset temperature for the formation of the Kondo groundstate \cite{Caldwell:2007iu}. We note that this onset temperature increases to about 100 K if we include the frequency range up to 30 meV, but in this case the agreement with the measured $\rho_\mathrm{DC}(T)$ is lost around 150 K.    

Before proceeding with further analysis we highlight a few observations that can be gleaned directly from the reflectivity data itself. With decreasing temperature several minima develop in the reflectivity spectra. An isobestic point (i.e. a frequency where the reflectivity is temperature independent) is visible at 0.13 eV. Both above and below this energy, minima start to develop as temperature is lowered. As we will see below, both minima roughly correspond to interband transitions. As temperature decreases below $\sim$ 60 K, a third minimum develops at the lowest measured energies (starting to become visible around 20 meV in the 50 K spectrum).  

Another feature that becomes more prominent with decreasing temperature is a phonon mode around 19.4 meV, as shown in the inset of Fig. \ref{Fig1}b. This mode has been observed for other hexaborides \cite{Kimura:1991ur, Degiorgi:1997tw, Ott:1997vw, Vonlanthen:2000ws, Werheit:2000er, Cho:2004gf, Perucchi:2004je, Kim:2005ii} and according to the symmetry of the crystal structure should be the $T_\mathrm{1u}$ mode. This mode has thus far not been seen in the IR spectra of SmB$_{6}$ \cite{Travalgilini:1984vy, Nanba:2002va, Gorshunov:1999hs, Hudakova:2004bh}, attesting to the high quality of our single crystal. A symmetry `forbidden' excitation was observed in recent Raman experiments \cite{Nyhus:1995jq,Valentine:2016wb} with a similar energy as our phonon mode and we suggest that the mode observed in the Raman spectra could thus be interpreted as part of the IR optical phonon branch observed here. We note that the mode becomes more prominent in the reflectivity spectra at low temperature due to `unscreening'. Specifically, the oscillator strength associated with this mode remains more or less temperature independent, but with decreasing temperature the free charge carrier density is reduced resulting in less effective screening of the mode. Finally, we note a weak but noticeable Fano-like asymmetry for this feature that becomes more prominent as the hybridization gap opens, similar to what has been observed in FeSi \cite{Damascelli:1997vd}.

\section{Results}\label{results_sec}
Next we turn our attention to the complex optical response functions. These were obtained from the reflectivity data using a Kramers-Kronig consistent variational routine \cite{Kuzmenko:2005jh}. When applied to reflectivity data, this approach is equivalent to a Kramers-Kronig transformation, but with slightly modified low and high frequency extrapolations (see appendix \ref{extrap}). We obtain all relevant optical quantities such as the complex optical conductivity $\sigma(\omega) = \sigma_1(\omega) + \eye \sigma_2(\omega)$ and dielectric function $\epsilon(\omega) = \epsilon_1(\omega) + \eye \epsilon_2(\omega)$ from the resulting variational dielectric function model.  

\subsection{Complex optical conductivity}\label{opt_cond}  
\begin{figure}[h]
\includegraphics[width = 8.6 cm]{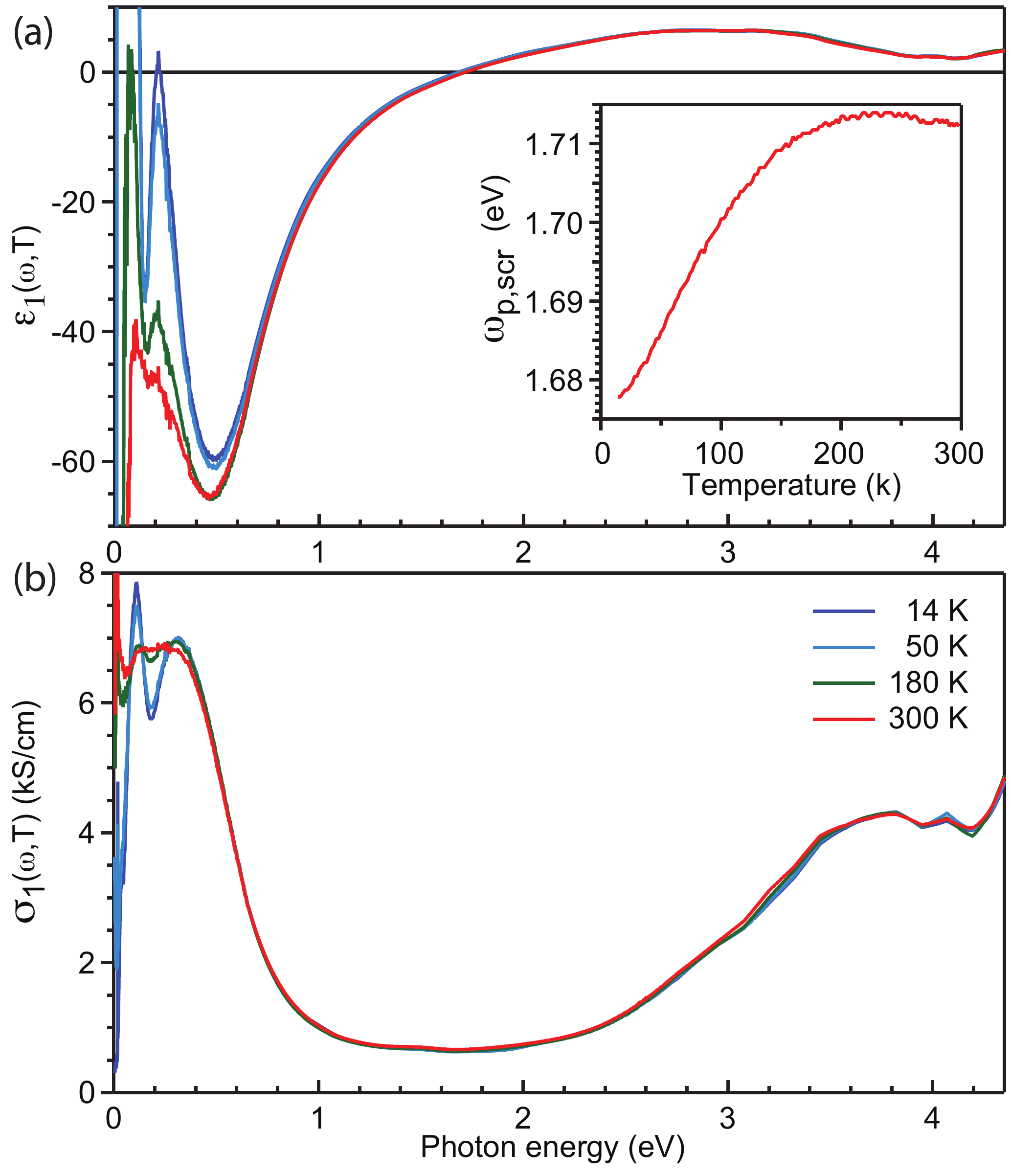}
\caption{(a): Dielectric function for selected temperatures (indicated in panel (b)) over the entire frequency range. At room temperature (red) the dielectric function starts out negative, indicative of a free charge response. As temperature decreases interband transitions start to dominate the low energy response. A zero crossing around 1.7 eV, corresponding to a transverse plasma oscillation, is present at all temperatures. The inset shows the temperature dependence of this zero crossing. (b): Optical conductivity at temperatures indicated. The optical conductivity displays several low energy interband transitions as well as weak structures around 2 eV. Above 2 eV a prominent series of interband transitions appear.}
\label{Fig2}
\end{figure}

Figure \ref{Fig2} presents the real part of the optical constants of SmB$_{6}$ over the entire measured frequency range. In panel \ref{Fig2}a we show the dielectric function $\varepsilon_{1}(\omega,T)$ and in panel \ref{Fig2}b the optical conductivity $\sigma_{1}(\omega,T)$ for the same temperatures as the reflectivity data in Fig. \ref{Fig1}. At room temperature, the dielectric function resembles that of a typical metal: negative at low energy and, with the exception of a structure around 125 meV, monotonically increasing as a function of energy. We observe a zero crossing around 1.7 eV at all temperatures, corresponding to the screened plasma frequency, $\omega^{2}_\mathrm{p,scr}(T)\equiv\omega^{2}_\mathrm{p}(T)/\varepsilon_{\infty}$. The inset of panel \ref{Fig2}a shows that $\omega_\mathrm{p,scr}$ decreases with temperature. This decrease is gradual at higher temperatures, but it starts to accelerate below roughly 200 K. We note that at lower temperatures the dielectric function displays several additional zero crossings, which we will discuss further below. 

At photon energies larger than $\omega_\mathrm{p,scr}$, $\varepsilon_{1}(\omega,T)$ remains positive with structures up until the highest measured frequencies. In this range (3 - 4 eV), the optical conductivity $\sigma_{1}(\omega,T)$ (Fig. \ref{Fig2}b) shows a strong interband transition. The origin of this transition is not entirely clear. According to LSDA+$U$ calculations it can have a different interpretation depending on the assumed Sm valence \cite{Antonov:2002uw}. If a divalent Sm$^{2+}$ configuration is assumed, the transition involves mostly B 2$p$ $\to$ Sm 5$d$. Alternatively, it corresponds to a mixture of B 2$p$ $\to$ Sm 5$d$ and Sm 5$d$ $\to$ Sm 4$f$ transitions in the case of trivalent Sm$^{3+}$. We further observe two weak structures in the optical conductivity around 1.5 and 2 eV. These transitions are not explicitly described in Ref. [\onlinecite{Antonov:2002uw}], but the calculated optical conductivity for the Sm$^{3+}$ configuration \textit{does} show a sharp structure in this range. They seem to originate from transitions between Sm 5$d$ $\to$ Sm 5$d$-4$f_{7/2}$ states and are exclusive to the trivalent Sm$^{3+}$ configuration. Our data therefore support the presence of some Sm$^{3+}$ in the system.

\begin{figure}
\includegraphics[width = 8.6 cm]{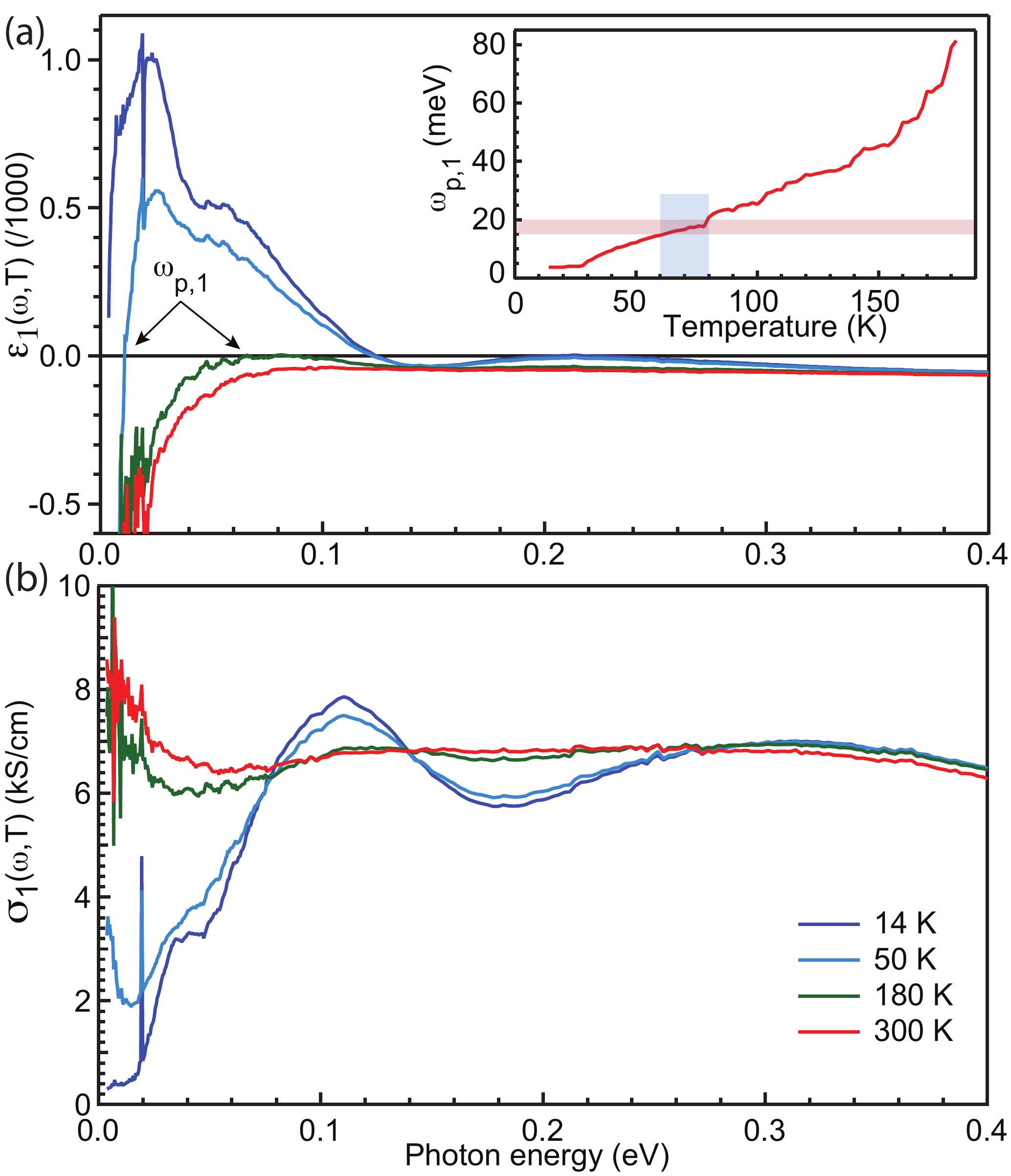}
\caption{(a): $\varepsilon_{1}(\omega,T)$ at selected temperatures (note the rescaling of the vertical axis by a factor of 1000.). At high temperatures $\varepsilon(\omega,T)$ is negative over the entire range shown. At 180 K a new zero crossing, labelled $\omega_\mathrm{p,1}$, occurs around 80 meV. As the inset shows, $\omega_\mathrm{p,1}$ shifts to lower energy with decreasing temperature and $\varepsilon_{1}(\omega,T)$ becomes positive over an extended range of photon energies. The red shaded area indicates the approximate energy of the optical gap edge and the blue area indicates the temperature range where $\omega_\mathrm{p,1}$ drops below this edge. (b): The corresponding $\sigma_{1}(\omega,T)$. We observe three distinct interband transitions (40, 120 and 300 meV) at the lowest temperatures. Note also the narrow Drude peak that is still clearly visible at 50 K.}
\label{Fig3}
\end{figure}

Figure \ref{Fig3} presents a more detailed view on $\varepsilon_{1}(\omega,T)$ and $\sigma_{1}(\omega,T)$ at low photon energy, which allows for a more careful examination of the temperature dependence in this range.  At room temperature the dielectric function in this photon energy range is always negative; however, as temperature is decreased below 180 K, a new zero-crossing appears, labeled $\omega_\mathrm{p,1}$. If we decrease temperature further, $\varepsilon_{1}(\omega,T)$ becomes positive over an extended range.  Figure \ref{Fig3}b shows that for temperatures between room temperature and 180 K the main change in the optical conductivity is a reduction in the spectral weight of the Drude peak. We will discuss the detailed evolution of the Drude spectral weight in the next section. Here we would like to point out that the Drude peak has been completely suppressed at 14 K pointing to the formation of a gap in the energy spectrum at the Fermi level. Below 180 K two transitions become more prominently visible in $\sigma_{1}(\omega,T)$: one around 0.11 eV and another around 0.32 eV, and both increasing in spectral weight as temperature decreases. At much lower temperatures a third transition becomes evident in $\sigma_{1}(\omega,T)$ around 0.04 eV. 

There are several possible origins for these three peaks. The LSDA+$U$ calculations of Ref. [\onlinecite{Antonov:2002uw}] again predict several interband transitions in the photon energy range below 1 eV, depending on the Sm valence. However, several aspects of this interpretation should be noted: First, the experimental transition around 0.3 eV seems to stem mostly from occupied, mixed $d$-$f$ $\to$ 5$d$-states for the Sm$^{2+}$ configuration, but it may also contain a component related to Sm 5$d$ $\to$ 4$f$ transitions for the Sm$^{3+}$ configuration. Compared to the experimental data reported in Ref.  [\onlinecite{Nanba:2002va}], the centre of this transition is somewhat lower in energy in our case. If this transition indeed has contributions from both Sm$^{2+}$ and Sm$^{3+}$, this reduction in the centre of the transition could be understood as arising from a different average valence and indicating a larger Sm$^{2+}$ component for our experiments. This hypothesis is further corroborated by the transition that we observe around 0.11 eV. This transition is also present in the experimental data of Ref.'s [\onlinecite{Gorshunov:1999hs}] and [\onlinecite{Nanba:2002va}], but in those experiments the transition is much weaker relative to the transition at 0.3 eV, whereas in our case it is much more prominent.  According to Ref. [\onlinecite{Antonov:2002uw}], this transition is between occupied, mixed $d-f$ states and unoccupied Sm 5$d$ states that are exclusive to the Sm$^{2+}$ configuration. The second aspect of note is that Antonov {\em et al.} predict that the lowest lying interband transitions occur around 50 meV and are between various hybridized Sm 5$d$ bands exclusive to the Sm$^{3+}$ configuration. This matches well with the structure that we see evolving with temperature just above the optical gap. As might be expected, the lowest energy transitions are most strongly affected as function of temperature as these bands involve transitions directly between hybridizing bands.  

We conclude this section with some remarks concerning the appearance of the additional zero-crossings in $\epsilon_1(\omega,T)$ at low temperature, and in particular the low energy crossing labeled $\omega_\mathrm{p,1}$. As the inset of panel \ref{Fig3}a shows, this crossing first appears below 180 K and quickly shifts to lower energy with decreasing temperature. It disappears below the lower limit of our experimental range below $T \lesssim 25$ K, but we are still able to infer its approximate energy position from our Drude-Lorentz model. The low temperature value, $\omega_\mathrm{p,1}$(18 K) = 3.5 meV, agrees reasonably well with the zero crossing reported in Gorshunov {\em et al.} (Ref. [\onlinecite{Gorshunov:1999hs}]). Although the temperature dependence of $\omega_\mathrm{p,1}$ does not show signs of changes at a particular temperature, a specific temperature scale can be defined as follows. With decreasing temperature the low energy spectral weight starts to decrease and an optical gap begins to form. As Fig. \ref{Fig3}b shows, the hybridization gap appears to have completely formed around 50 K (based on the appearance of the low energy peak around 0.04 eV). We estimate that the optical gap is approximately 15 to 20 meV wide, based on the 14 K spectrum and using the onset of absorption just below the phonon mode as a measure. In the inset of Fig. \ref{Fig3}a we indicate the approximate size of the optical gap as the red shaded area and note that the zero crossing in $\varepsilon_{1}(\omega,T)$ falls below the gap edge at low temperature. The temperature range where this happens is indicated by the blue shaded area and spans the temperature range 60 - 80 K.
 
\begin{figure*}
\includegraphics[width=17.8cm]{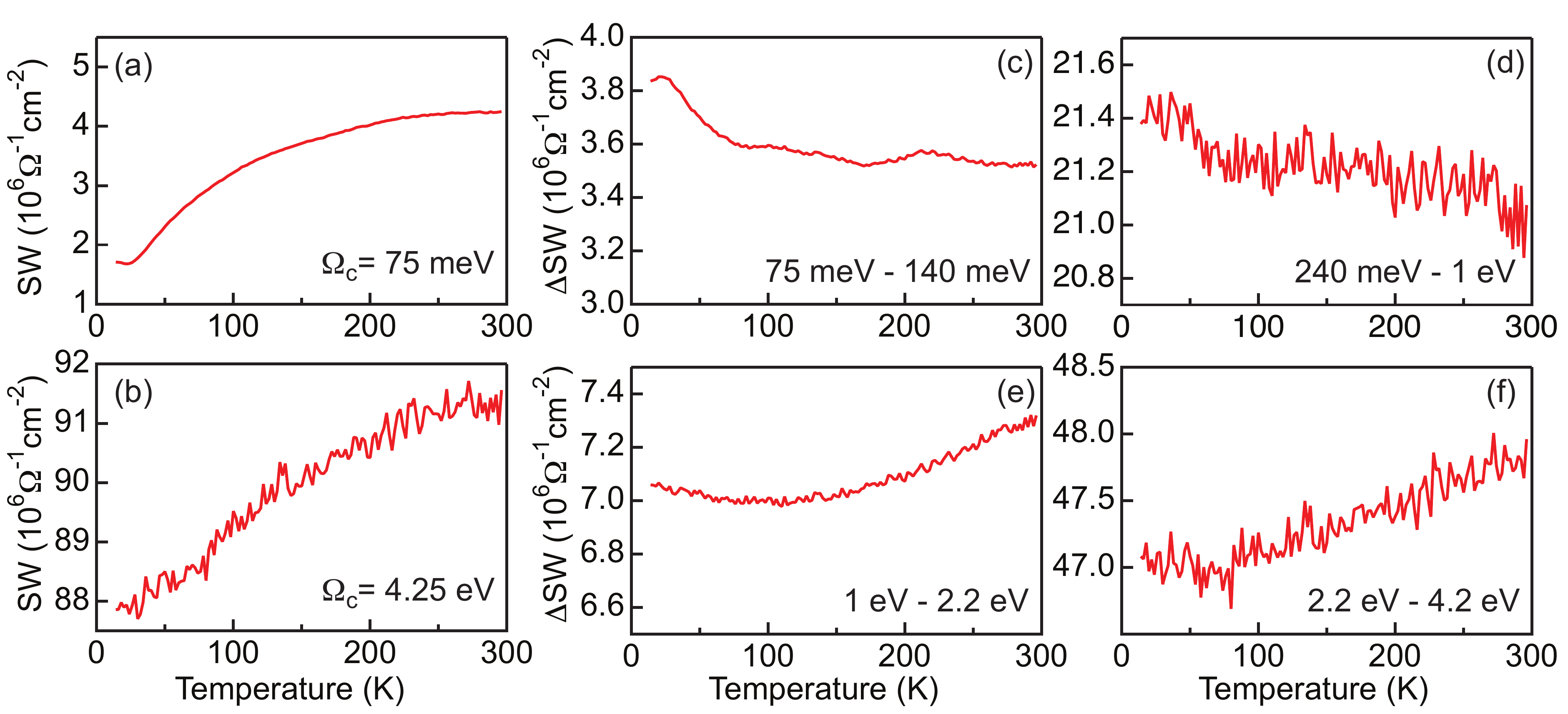}
\caption{(a,b): Spectral weight as a function of temperature, integrated up to a cutoff $\Omega_{c}$ indicated in each panel. The spectral weight below 75 meV is lost, which is not recovered in the full measured experimental range. Note that the vertical axes of the two panels are chosen such that the relative change is the same for both ($4.5\cdot10^{6}$ $\Omega^{-1}$cm$^{-2}$). (c-f): Integrated spectral weight as function of temperature between two cutoffs indicated in each panel. One thing to note is that in each panel a small amount of spectral weight increase occurs below approximately 70 K. Further details are discussed in the text.}
\label{FigSW}
\end{figure*}

\subsection{Spectral weight transfer}\label{spec_transf}
Having identified the main features of the optical response functions, we now turn our attention to the somewhat more subtle changes in their temperature evolution by presenting an analysis of spectral weight transfers associated with the destruction of the metallic state and the formation of the hybridization gap.  As temperature decreases the hybridization of the $d$ and $f$ states is expected to result in the formation of an energy gap near (or at) the Fermi level \cite{Martin:1979fx}. The formation of this state takes place at the cost of mobile $d$-electron states in favor of a larger occupation of $f$-electron states and a reduction of the Drude spectral weight results. This can be quantified by examining the partial integrated spectral weight, 
\begin{equation}\nonumber      
\mathrm{SW}(\Omega_{a},\Omega_{b},T)=\int_{\Omega_{a}}^{\Omega_{b}}\sigma_{1}(\omega,T)d\omega.
\end{equation}
When applied to materials where the optical properties change as function of an external parameter, the integral is often restricted to finite frequency intervals in order to detect associated transfers of spectral weight. The energy range over which these transfers take place can provide information on the interactions involved in the transition \cite{Katsufuji:1995a,Rozenberg:1996,Molegraaf:2002df}. 

Inspection of Fig. \ref{Fig2} and Fig. \ref{Fig3}, allows us to anticipate approximately the relevant energy ranges. The low energy range is defined by $\Omega_{a}=$ 0 meV and $\Omega_{b}\approx$ 75 meV and the temperature dependence of the $\mathrm{SW}$ in this range is shown in Fig.  \ref{FigSW}a. The integrated spectral weight $\mathrm{SW}(75~\mathrm{meV},T)$  continuously decreases as the temperature is lowered. Assuming that this depletion can be ascribed entirely to a collapse of the Drude peak, we use $\Delta \mathrm{SW}=\mathrm{SW}(300~\mathrm{K})-\mathrm{SW}(14~\mathrm{K})$ as an estimate of the free carrier density at room temperature. From Fig.  \ref{FigSW}a we obtain $\omega_\mathrm{p}\sim\sqrt{120\Delta \mathrm{SW}/\pi}\approx$ 9850 cm$^{-1}$, which corresponds to $n_\mathrm{free}\approx 1.08\pm 0.08\cdot10^{21}$ cm$^{-3}$ assuming that there is no mass renormalization (e.g. assuming $m_\mathrm{b}\approx1$ at room temperature). This value is in good agreement with early estimates of the room temperature carrier density \cite{JWAllen:1979a}. The error bar on this number was determined using the method presented in appendix \ref{SW_tr}. If we express this in terms of carriers per formula unit, we find that the loss of spectral weight with decreasing temperature corresponds to 0.076 carriers per SmB$_{6}$ unit. This reduction is in close agreement with a reduction of the Sm valence observed in temperature dependent x-ray absorption measurements \cite{Mizumaki:2009ju}, where the estimated average valence of the Sm ions changes from 2.58 at room temperature to 2.5 at low temperature.

One expects that the Drude weight decreases in a simple semiconductor with the chemical potential inside the band gap (or very close to the bottom/top of a band) as a result of changes in the available phase space for scattering and the thermal distribution of free carriers. In SmB$_{6}$ at intermediate temperatures the chemical potential sits very close to the hybridization zone between the mobile $d$ band and the localized $f$-levels, while at low temperature the hybridization gap opens at the Fermi level as evidenced by the optical gap. The loss of spectral weight could therefore result from a similar kinematic effect as that occurring in semiconductors.  In appendix \ref{app_tb} we calculate the temperature dependence of the optical constants for such a scenario. 

Applying the same analysis of the spectral weight to the calculated optical conductivity shows that the Drude spectral weight is transferred in the energy range just above the optical gap such that the total spectral weight remains constant. This is not the case for SmB$_{6}$ as Fig. \ref{FigSW}b shows. If we integrate the optical conductivity over the entire range measured, we find that the total integrated spectral weight decreases with about the same amount. This indicates that a significant amount of Drude spectral weight is being redistributed to energies well above $\sim 4.35$ eV. Moreover, this effect is well beyond what is expected due to simple kinematic effects associated with the opening of the hybridization gap. This situation is somewhat reminiscent of spectral weight transfers taking place in the cuprate high T$_{c}$ superconductors \cite{Molegraaf:2002df}. In the cuprates spectral weight is transferred from low energy to the scale of the effective Coulomb interaction ($\approx$ 2-3 eV in the cuprates) as the superconducting gap opens.  For the case at hand, the on-site $4f$ Coulomb interaction was estimated to be $U_\mathrm{eff}\approx$ 7 eV \cite{Antonov:2002uw}. This is beyond our experimental window; however, a previous optical experiment by Kimura {\em et al.} \cite{Kimura:1993vi} indeed shows interband transitions around 5 and 10 eV. 

Our spectral weight analysis suggests that the hybridization between $d$ and $f$ states involves energy scales on the order of the effective Coulomb interaction. Surprisingly, almost all of the Drude spectral weight lost in the formation of the Kondo insulating state is transferred to this high energy scale.  The stark contrast between the spectral weight redistribution in the experiment and the tight-binding model perhaps serves as an illustration of the idea \cite{Martin:1979fx} that the Kondo insulator can be understood as the large $U$ version of a hybridized band insulator. The effective Kondo interaction $J\propto \Delta^{2}/U$ becomes small in the large $U$ limit. Transfers of spectral weight involving large energy ranges have been observed in several other Kondo insulators \cite{Bucher:1994if, Schlesinger:1993bs}, which therefore seems a generic feature associated with the formation of the Kondo insulating state. The loss of spectral weight can therefore be seen as a signature of strong electron-electron interactions and the energy scale over which it is recovered as a measure of the effective Coulomb interaction.   

The remaining panels of Fig. \ref{FigSW} demonstrate that the overall reduction of spectral weight is masking more subtle shifts in spectral weight. For example, the spectral weight of the interband transition around 0.11 eV increases at low temperatures, as is clearly visible in Fig. \ref{Fig3}b. To estimate if this is the pile-up of spectral weight just above the gap edge anticipated above, we calculate the partial sum rule, $\Delta\mathrm{SW}(T)$, integrated from 75 meV to 140 meV (Fig. \ref{FigSW}c). Here we observe a different trend in spectral weight as function of temperature: the spectral weight contained in this transition remains more or less constant as function of temperature down to approximately 60 - 70 K, where the spectral weight suddenly begins to increase. This temperature is close to the temperature where the reflectivity starts to deviate from Hagen-Rubens behavior (see Fig. \ref{Fig1}a), suggesting that the enhancement of this transition is indeed related to the opening of the hybridization gap. Note, however, the difference in SW scales. The total Drude spectral weight lost is $2.55\cdot10^{6}$ $\Omega^{-1}$cm$^{-2}$, while the increase of spectral weight around 0.11 eV is of the order of $0.2\cdot10^{6}$ $\Omega^{-1}$cm$^{-2}$ (panel \ref{FigSW}c). We find a similar increase in the range between 0.24 eV and 1 eV (panel \ref{FigSW}d). $\Delta\mathrm{SW}(T)$ between 1 eV and 2.2 eV (Fig. \ref{FigSW}e) is relatively weak, with a change in slope observed as function of temperature. Finally, panel, \ref{FigSW}f, shows $\Delta\mathrm{SW}(T)$ in the range between 2.2 and 4.2 eV. The spectral weight in this transition mostly decreases with temperature, but a similar change in slope can be observed around 60 - 70 K, resulting in a small increase of spectral weight of again $0.2\cdot10^{6}$ $\Omega^{-1}$cm$^{-2}$. To summarize, the total spectral weight decreases, but we observe a small increase in spectral weight in all of the \emph{interband} transitions as the Kondo insulating state forms. In appendix \ref{SW_tr} we show that this result is robust against substantial shifts of the experimental reflectivity data and corresponding changes in the spectral weight analysis.      

As discussed above, the interband transitions at 0.11 eV and 0.32 eV contain a significant component originating from mixed $d-f$ $\to$ 5$d$ transitions. The increase in spectral weight below 60 - 70 K can therefore possibly be linked to changes taking place in the hybridization of the $d$ and $f$ states. In recent ARPES works the temperature dependence of the $d-f$ hybridization was investigated in detail \cite{Min:2014kp,Denlinger:2013tk}. It has been observed that one of the $f$-levels shifts from above (low temperature) to below (high temperature) the Fermi level. The temperature where this state crosses the Fermi level is approximately 60 K, very close to the temperature where we observe changes taking place in (i) the low energy reflectivity and (ii) the temperature dependence of the integrated spectral weight. It is possible that the small feature at 0.04 eV just above the gap edge (see Fig. \ref{Fig3}b) arises from interband transitions between these occupied and unoccupied mixed $d-f$ states, as the energy matches reasonably well with the splitting observed in Ref.  [\onlinecite{Denlinger:2013tk}]. Comparing to our toy-model calculation, this then implies a hybridization parameter of approximately 20 meV. These authors also notice a second effect: namely a loss of the `coherency' of the $f$-states with increasing temperature. This change in coherence (observed as changes in the 4$f$ - amplitude and width in the ARPES spectra) is a gradual trend however and not particularly linked to an onset temperature\cite{Denlinger:2013tk}. It is likely that these changes in coherence are reflected in the smearing of spectral features in the optical spectra. 

\section{Summary}\label{summary}
We have investigated the temperature dependent optical properties of SmB$_{6}$ in detail. From the reflectivity data we estimate that the high temperature metallic state is destroyed below 60 - 70 K. A new feature observed in the reflectivity data is a phonon mode with an energy of 19.4 meV, which is related to the T$_\mathrm{1u}$ mode associated with the rattling of the Sm ion against the boron cages. An analysis of the optical spectra shows that the destruction of the metallic state is a gradual trend, with an approximate onset temperature around 200 K. A comparison of the measured interband transitions and LSDA+$U$ calculations indicates the presence of ions with varying valence in this material. The destruction of the metallic state is accompanied by a loss of low energy (Drude) spectral weight that is not recovered in the experimental range measured. In contrast, a representative tight-binding model calculation shows that this spectral weight should be recovered on an energy scale corresponding to the hybridization strength of the $d$ and $f$-states. Our analysis suggests that this spectral weight is instead shifted over an energy range involving the effective Coulomb interaction ($U_\mathrm{eff}\approx$ 7 eV). We suggest that this is a signature of the important role played by strong electron-electron interactions in this material.

\section{Acknowledgement}
EvH would like to thank S.V. Ramankutty, M.S. Golden, Yu Pan and A. de Visser for fruitful discussions and the IoP for support. S.J. was supported by the University of Tennessee Science Alliance JDRD program; a collaboration with Oak Ridge National Laboratory

\appendix
\section{Reflectivity extrapolations}\label{extrap}
The traditional method for obtaining the complex optical conductivity or dielectric function from reflectivity data is to make use of a Kramers-Kronig transformation. In such an approach the phase angle of the complex reflection coefficient is calculated by making use of the Kramers-Kronig relation to the measured reflectivity. In this case, extrapolations to low (zero) and high (infinite) frequency have to be made beyond the experimental range over which the reflectivity has been measured. Typical expressions for these extrapolations are the Hagen-Rubens approximation as discussed in section II for the low frequency extrapolation and these extrapolations always introduce some uncertainty in the resulting optical quantities. In our analysis we make use of a different approach that has been described in ref. [\onlinecite{Kuzmenko:2005jh}]. When applied to reflectivity data alone, this approach is equivalent to a Kramers-Kronig transformation, but with slightly different extrapolations for low and high frequency.    

The first step in this procedure is to fit the reflectivity spectrum at a given temperature with a Drude-Lorentz model. 
\begin{figure}
\includegraphics[width = 7 cm]{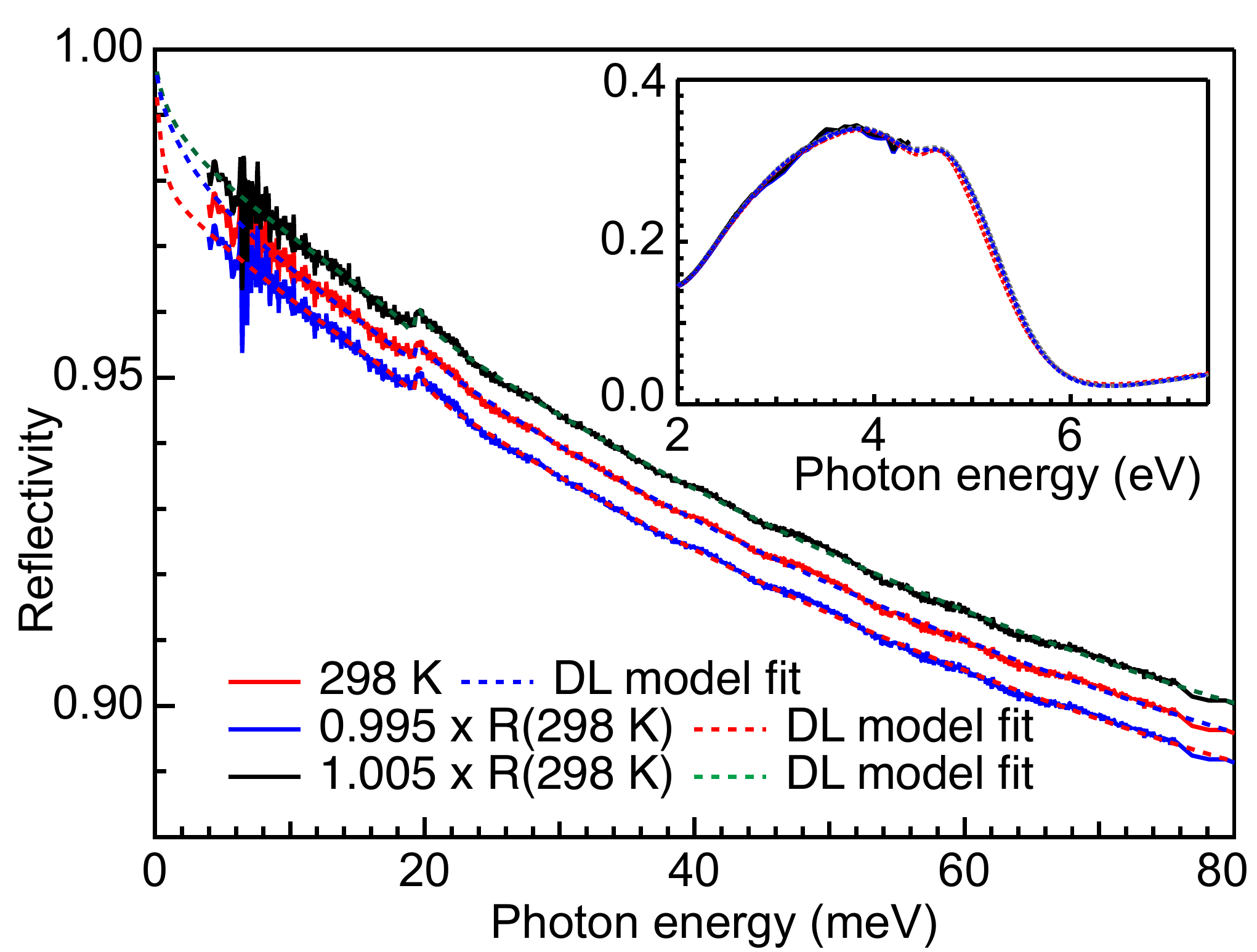}
\caption{Room temperature reflectivity data and the fit (red line and blue dashed line respectively) calculated from the Drude-Lorentz model at 298 K. Also shown are the scaled reflectivity data and fits used to make the error bar estimate of the spectral weight transfer (see appendix \ref{SW_tr}). The inset show the high frequency part of the spectrum and the extrapolation resulting from the Drude-Lorentz model.}
\label{FigRex}
\end{figure}
An example of a fit of the room temperature and the low and high frequency 'extrapolations' are shown in Fig. \ref{FigRex}. The fitted curve shown is still without the variational function added on top of it. Note that the Drude-Lorentz model already provides a very accurate description of the reflectivity data. 
The second step in this procedure is to add a variational dielectric function on top of this model to reproduce all the detailed features (including noise) of the reflectivity spectrum. The full Drude-Lorentz + variational dielectric function model then provides full access to any of the optical quantities of interest. 

There are two major advantages to this approach. The first is that there are no artifacts at the boundaries of the experimental data: the flexibility of the model allows for a smooth transition between extrapolation and experimental data. The second advantage is that some of the parameters are more robustly determined. For example, the value of $\varepsilon_{\infty}$ is determined by the reflectivity data over a very broad energy range. 

\section{Tightbinding Model}\label{app_tb}
In order to examine the kinematic effects we considered a simple two-band tight-binding model designed to reproduce the salient features of the SmB$_{6}$ optical spectra. The Hamiltonian is  
\begin{eqnarray}\nonumber
H&=&\sum\limits_{\mathbf{k},n,\sigma }   \epsilon^\pdag_{n} (\mathbf{k}) c_{n \mathbf{k} \sigma } ^{\dagger} c^\pdag_{n \mathbf{k} \sigma  }    + \sum\limits_{\mathbf{k},n,m,\sigma    }  t^\pdag_{nm} (\mathbf{k}) c_{n \mathbf{k} \sigma }^{\dagger} c^\pdag_{m \mathbf{k} \sigma  } \\ 
&=&\sum\limits_{\mathbf{k},n,m,\sigma } c_{n \mathbf{k} \sigma } ^{\dagger} H^\pdag_{nm}(\mathbf{k})c_{m \mathbf{k} \sigma }^{\pdag} .
\end{eqnarray}
Here, $\epsilon_n(\mathbf{k})$ is the band dispersion of the $n^{\text{th}}$ band; $c^\dagger_{n \mathbf{k},\sigma}$ ($c^\pdag_{n \mathbf{k},\sigma}$) is the  electron creation (annihilation) operators for the $n^{\text{th}}$ band; $t_{nm}(\mathbf{k})$ is the hybridization parameter between $n^{\text{th}}$ and $m^{\text{th}}$ bands; and $\sigma$ is the spin of the electron. Our two-band model then consists of a dispersive band $\epsilon_1(\mathbf{k}) = -2t_{d}[\cos(k_xa) + \cos(k_ya) + \cos(k_za)] - \mu_0$, and a dispersionless (localized) band $\epsilon_2(\mathbf{k}) = \epsilon_2$. The hybridization parameter between the two bands is momentum independent with $t_{12} = t_{21} = \Delta$. We then set ($t_{d},~\mu_{0},~\Delta$) = ($0.5,~-1,~-0.1$) eV and $\epsilon_2$ = -15 meV, such that the band structure mimics approximately the band dispersion around the X-point of the Brillouin zone \cite{Lu:2013wc}. These bands hybridize with a hybridization parameter $\Delta=0.1$ eV, resulting in a direct optical gap of 0.2 eV and an indirect gap of approximately 20 meV. We choose the hybridization parameter to be temperature independent as this is sufficient to demonstrate the energy range over which spectral weight is redistributed. 

\begin{figure}
\includegraphics[width = 7 cm]{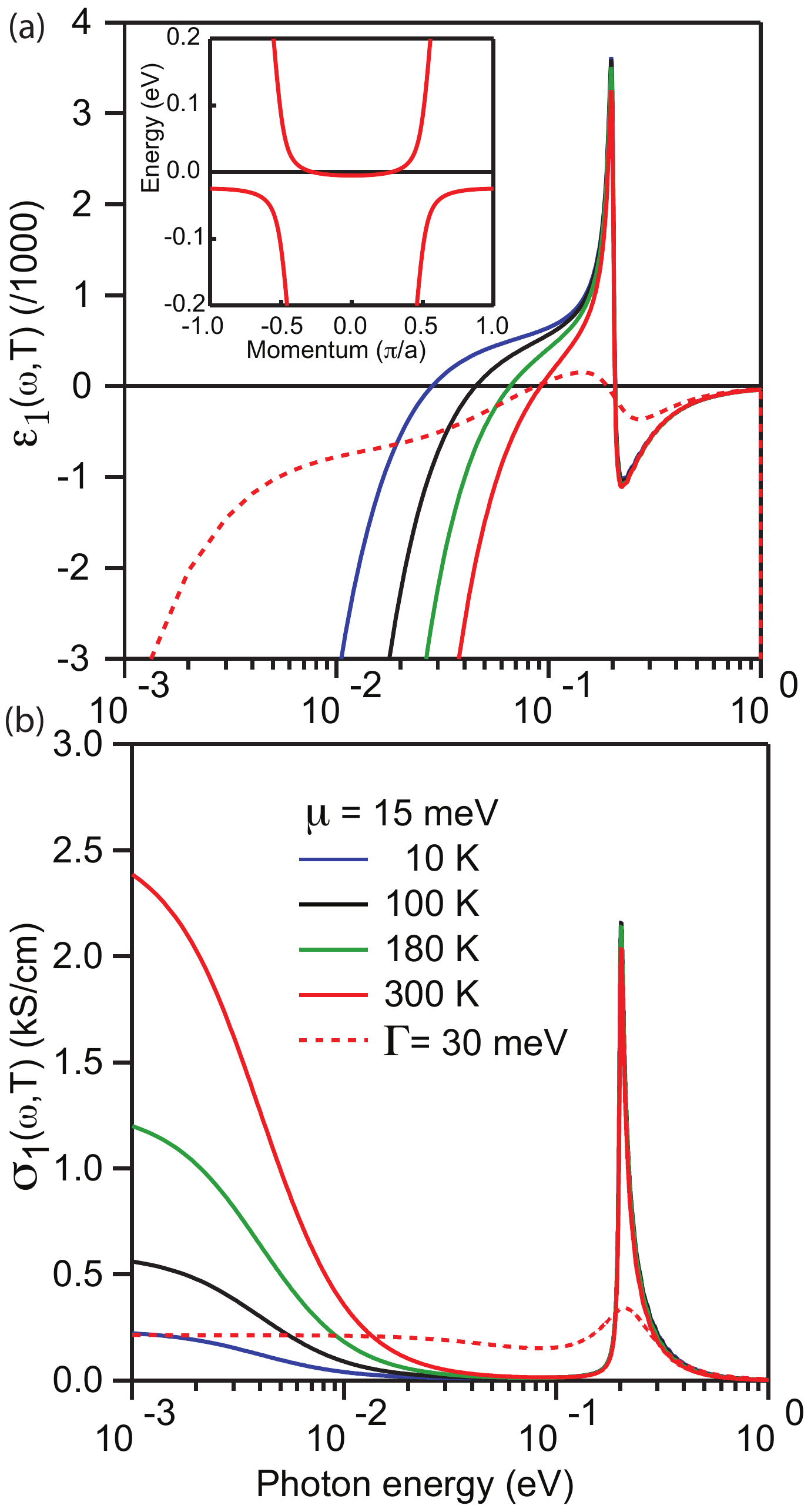}
\caption{(a): Calculated real part of the dielectric function at selected temperatures, displaying two zero crossings (one at energy just above 1 eV). The inset shows the hybridized band model used in the calculation, plotted along the $\vec{k}=(k_{x}, 0, \pi/2)$ direction. The chemical potential sits 15 meV above the band bottom. (b): optical conductivity showing the reduction of the Drude peak with decreasing temperature. The spectral weight is transferred to the interband transition around 0.2 eV. The dashed curve in both panels shows the 300 K spectrum calculated with a large impurity scattering rate.}
\label{FigTH}
\end{figure}

The optical conductivity $\sigma_{i,j}(\mathbf{q},\omega)$ is evaluated from the Kubo formula, in which the induced electric current density $J$ is approximated to be linear to the external field $E$ as $J_i(\mathbf{q},\omega) = \sum_j \sigma_{i,j}(\mathbf{q},\omega) E_j(\mathbf{q},\omega)$, where $i, j$ are the cartesian directions, $\sigma_{i,j}(\mathbf{q},\omega)$ is the conductivity tensor; and $\mathbf{q}$ and $\omega$ are the wavevector and energy of the applied field \cite{mahan_book}. For optics, we take the $\sigma_{ij}(\mathbf{q}\rightarrow 0,\omega) \equiv \sigma_{ij}(\omega)$ limit. The detailed derivation of $\sigma_{ij}(\omega)$, including multiband effects, can be found in Ref. [\onlinecite{Atkinson:1995ej}]. The real part of $\sigma_{ij}(\omega)$ is given
by 
\begin{eqnarray}\nonumber
\text{Re}[\sigma_{ij}(\omega)]&=&\frac{e^2\hbar}{2\pi V} \sum\limits_{\mathbf{k}} \int_{-\infty}^{\infty} dz 
\frac{n_f(z)-n_f(z+\omega)}{\omega} dz \\ 
&\times&
\text{Tr}[A(\mathbf{k},z) \eta_i(\mathbf{k})  A(\mathbf{k},z+\omega)  \eta_j(\mathbf{k})],  
\end{eqnarray}
where $A_{nm}(\mathbf{k},z)$ is the imaginary part of the electron Greens function $G_{nm}^{-1}(\mathbf{k},z) = (z +\eye \Gamma)\delta_{nm} - H_{nm}(\mathbf{k})$, $\eta_i(\mathbf{k}) = \frac{1}{\hbar} \frac{\partial}{\partial k_i} H_{nm}(\mathbf{k}) $ is the vertex function, $n_f(z)$ is the Fermi-Dirac function, and $V$ is the volume of the crystal. Here, $\Gamma = 2$ meV is a phenomenological broadening introduced by impurity scattering. The dielectric function is related to the conductivity by $\epsilon(\omega) = 1 + 4\pi \eye \sigma(\omega)/\omega$, where $\sigma(\omega)$ is the complex-valued optical conductivity with the imaginary part obtained from the Kramers-Kronig relations.  

The calculated optical spectrum is shown in Fig. \ref{FigTH}. Figure \ref{FigTH}a shows the calculated dielectric function at selected temperatures. It captures the main features of the experimental spectrum: at low frequency there is a zero-crossing followed by an interband transition and then two more zero crossings (the zero crossing corresponding to the screened plasma frequency sits outside the calculated window). The model also captures the main temperature dependencies of the experimental data with the exception of the disappearance of the low frequency zero-crossing ($\omega_\mathrm{p,1}$ in the inset of Fig. 3a). Another difference is the broadening of the interband transitions with increasing temperature. This may be related to the temperature dependence of the scattering rate $\Gamma$, which we have neglected here. Figure \ref{FigTH}a also shows the optical spectrum calculated with an impurity broadening $\Gamma$ = 30 meV, providing a better agreement with the data at elevated temperature. Note that the increased broadening does not affect the energy of the zero-crossing.

The calculated optical conductivity is shown in Fig. \ref{FigTH}b. At high temperatures the results display a clear Drude peak as expected. With decreasing temperature, the Drude peak is gradually suppressed, which is associated with larger occupation of the band bottom. This effect is large enough that the resistivity increases by an order of magnitude as temperature is reduced. It is obvious that the integrated spectral weight below 0.1 eV is significantly reduced at low temperature. Our calculations show that within the model, this spectral weight is redistributed to higher energies, and most of it is transferred to the interband transition centered at 0.2 eV. However, a few percent of the spectral weight is redistributed over a wider energy range comparable to the full bandwidth of the model.

\section{Error estimate spectral weight transfer}\label{SW_tr}
The accurate estimation of the absolute value of the integrated spectral weight is complicated and to some extent depends on choices made in the conversion of the experimental data (e.g. reflectivity) to the complex optical quantities. In appendix \ref{extrap} we have already discussed the method used to extract the complex optical quantities. The purpose of this appendix is to show that the main conclusions of the article (i.e. a collapse of the Drude peak and the associated transfer of spectral weight over a large energy range to high energy) are a robust feature that do not depend significantly on the modeling choices made in our analysis. We will show that although the \textit{absolute} value of the spectral weight is very sensitive to the specific model chosen, the \textit{relative} value between different temperatures can be accurately determined.
\begin{figure}[t]
\includegraphics[width = 7 cm]{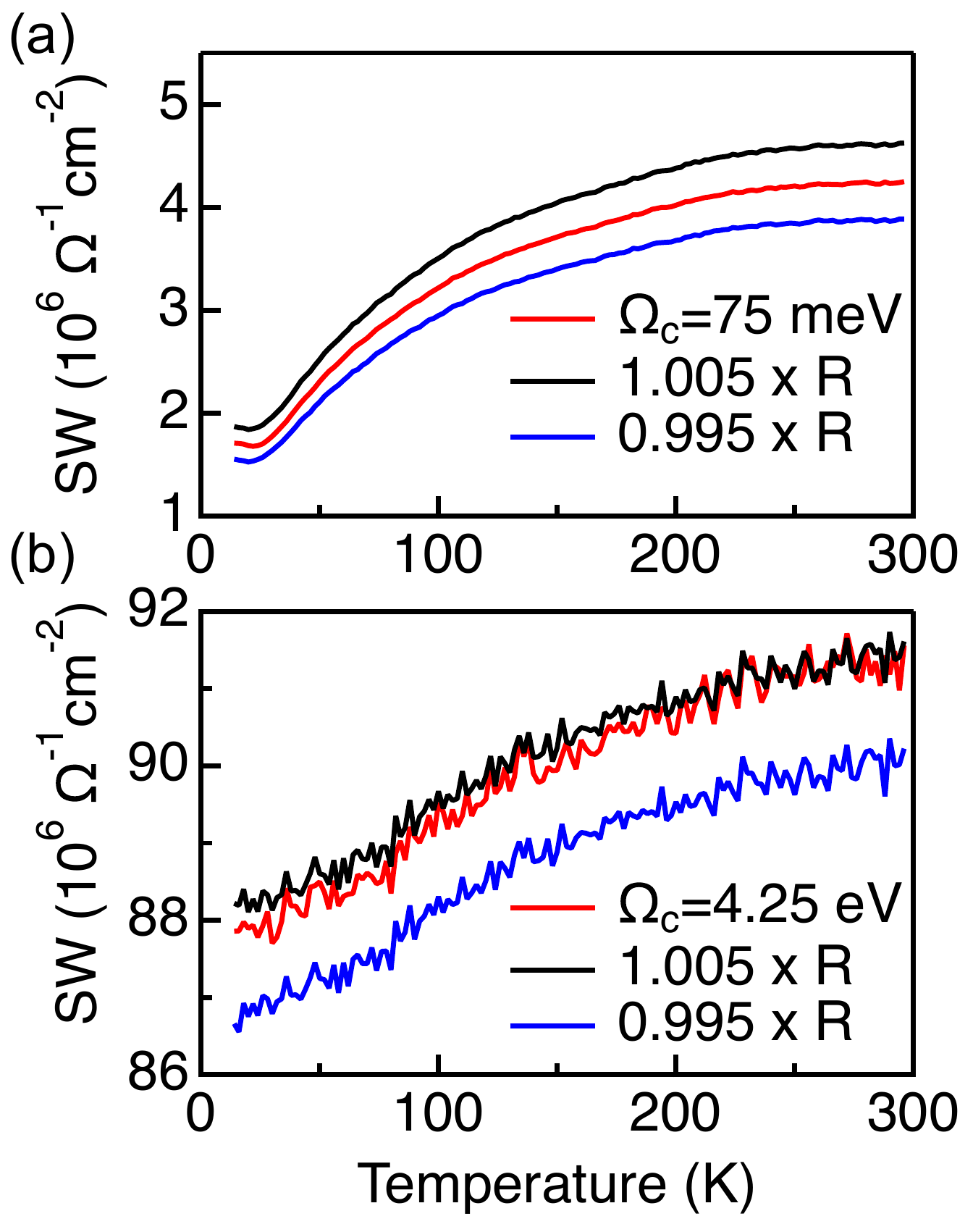}
\caption{(a): integrated spectral weight for a cutoff frequency of $\Omega_{c}$ = 75 meV. Shown are SW(T) as presented in Fig. \ref{FigSW} and SW(T) obtained when the data is scaled up (black) and down (blue) by 0.5 $\%$. (b): the same as for panel (a), but now with the cutoff frequency $\Omega_{c}$ = 4.25 eV.}
\label{Fig_SWerr}
\end{figure}

As a result of the highly detailed temperature dependence used to measure the reflectivity data over a very wide energy range, we can make a reasonable estimate of the error bar on our reflectivity spectrum. Even in the noisiest part of the spectrum (between 4 and 12 meV) the noise is only 0.4 $\%$ of the total reflectivity. The systematic error bar on the data (0.2 $\%$) can be estimated from the reproducibility of the many temperature sweeps performed (ranging from 2 cooling-warming cycles in the far infrared to up to 10 cycles in the UV range). To make an estimate of the error introduced in the determination of the spectral weight shifts we therefore follow the following procedure. We scale the reflectivity data up and down by 0.5 $\%$ (constant for all temperatures and well beyond the actual error) as shown in Fig. \ref{FigRex}. We then perform the full analysis to determine the spectral weight based on these scaled datasets. 
The results are summarized in Fig. \ref{Fig_SWerr}. As expected, the Drude spectral weight (Fig. \ref{Fig_SWerr}a) slightly increases if the reflectivity data is scaled upwards, while it decreases when the data is scaled downwards. The most important point however is that the \textit{relative} change between room temperature is qualitatively unchanged and quantitatively very similar. The change amounts to an error of about 8 $\%$ on the absolute value of the Drude weight. The same holds true if we calculate the integrated spectral weight up to a cutoff frequency $\Omega_{c}$ = 4.25 eV. The error bar on the relative change between low and high temperature is about 3$\%$, but most importantly the total spectral weight decreases irrespective of the shift of the reflectivity data. Comparing panels (a) and (b) of fig. \ref{Fig_SWerr}, we note that the redistribution of the additional Drude weight does not seem to result in an additional transfer of weight to energies beyond our experimental window. Instead this weight is already recovered in the interband transition between 3-4 eV.

\end{document}